\documentclass[twocolumn]{revtex4}
\usepackage{epsfig}
\begin{document}

\title{ Measuring carrier density in parallel conduction 
layers of quantum Hall systems}

\author{
M. Grayson, F. Fischer}

\affiliation{
Walter Schottky Institut, Technische Universit\"at M\"unchen, D-85748 
Garching, Germany\\}

\date{17 Nov 2004\\
Version 6}

\begin{abstract}
An experimental analysis for two parallel conducting
layers determines the full resistivity tensor of the parallel layer, 
at magnetic fields where the other layer is in the quantum Hall regime.  
In heterostructures which exhibit parallel conduction in the 
modulation-doped layer, this analysis quantitatively determines the 
charge density in the doping layer and can be used to estimate the 
mobility.  To illustrate one application, experimental data show 
magnetic freeze-out of parallel conduction in a modulation doped 
heterojunction.  As another example, the carrier density of a minimally 
populated second subband in a two-subband quantum well is determined.  A simple 
formula is derived that can estimate the carrier density in a highly resistive 
parallel layer from a single Hall measurement of the total system.

\end{abstract}

\maketitle

Various high-mobility heterostructures contain parallel conducting layers
whose individual conductance parameters need to be characterized.  Such
systems include double-quantum well systems and quantum wells with a
second occupied subband.  Another system of practical interest is a
high-mobility two-dimensional electron system (2DES) with a parallel
conducting modulation-dopant layer.  In all cases, the well-known
signature of parallel conduction is that the minima in the longitudinal
resistance $R_{xx}$ do not go to zero in the quantum Hall limit.  One
standard method for separately characterizing the layer densities is the
Shubnikov-de Haas (SdH) method, which requires that both systems show
oscillations in $R_{xx}$ in the same magnetic field range $B$.  These
oscillations are then Fourier transformed against $1/B$ to give the
densities of both systems \cite{Shayegan}.  However, this method cannot 
be used if one of
the systems has sufficiently low mobility that SdH oscillations are
indiscernable, and this method gives no separable measurement of $R_{xx}$
for each layer.  Other methods have been introduced which can determine
resistivities in non-quantized parallel channels \cite{Burgt,Hurd,Kane}, 
but to date an explicit treatment of the quantized case is lacking.  In 
this paper, we present a new analysis of magnetotransport data which 
determines both components of the resistivity tensor $\rho^{||}_{xx}$ and 
$\rho^{||}_{xy}$ in one layer, at values of the $B$ field where the {\em 
other} layer is in the quantized Hall regime.  The density of carriers in 
the parallel layer then follows directly $n^{||} = {B} /
{\rho^{||}_{xy}e}$, with $e$ the electron charge.

We wish to first find an expression for the resistivity tensor of the
combined quantum Hall and parallel conducting layers since resistivity is 
the experimentally measured quantity.  We define separate resistivity 
tensors for the quantum Hall and parallel conduction layers as:

\begin{equation}
\label{rhobare}
\begin{array}{ccc}
\bf{\rho^{Q}} = \left( \begin{array}{cc}
0 & \frac{h}{\nu e^2} \\
-\frac{h}{\nu e^2} & 0 
\end{array} \right),
&
&
\bf{\rho^{||}} = \left( \begin{array}{cc}
\rho^{||}_{xx} & \rho^{||}_{xy} \\
-\rho^{||}_{xy} & \rho^{||}_{xx} \\
\end{array} \right)
\end{array}
\end{equation}    

\noindent Assuming that the in-plane electric fields seen by the two
layers are equal, the corresponding conductivities for each layer,

\begin{equation}
\label{sigmabare}
\begin{array}{ccc}
\bf{\sigma^{Q}} = \left( \begin{array}{cc}
0 & -\frac{\nu e^2}{h} \\
\frac{\nu e^2}{h} & 0 
\end{array} \right),
&
&
\bf{\sigma^{||}} = \frac{1}{|\bf{\rho^{||}}|}
\left( \begin{array}{cc}
\rho^{||}_{xx} & -\rho^{||}_{xy} \\
\rho^{||}_{xy} & \rho^{||}_{xx} \\
\end{array} \right)
\end{array}
\end{equation}    

\noindent can be summed to yield the sheet conductivity of the combined
system 

\begin{equation}
\label{sigmtot}
\bf{\sigma^{tot}} = \bf{\sigma^{Q}} + \bf{\sigma^{||}} =
\frac{1}{|\rho^{||}|} 
\left( \begin{array}{cc}
\rho^{||}_{xx} & -\rho^{||}_{xy} - |\rho^{||}|\frac{\nu e^2}{h} \\
\rho^{||}_{xy} + |\rho^{||}|\frac{\nu e^2}{h} & \rho^{||}_{xx} \\
\end{array} \right)
\end{equation}

\noindent where $|\bf{\rho^{||}}| = \rm{det}(\bf{\rho^{||}})$.  Inverting
$\bf{\sigma^{tot}}$ gives the total resistivity tensor, $\bf{\rho^{tot}} =
(\bf{\sigma^{tot}})^{-1}$ whose components are directly measured in
experiment:

\begin{equation}
\begin{array}{ccc}
\label{rhotot}
\rho^{tot}_{xx} = \frac
{|\bf{\rho^{||}}| \rho^{||}_{xx}}
{{\rho^{||}_{xx}}^2 + (|\bf{\rho^{||}}|\frac{\nu e^2}{h} + \rho^{||}_{xy})^2},
&
&
\rho^{tot}_{xy} = \frac
{|\bf{\rho^{||}}| (|\bf{\rho^{||}}|\frac{\nu e^2}{h} + \rho^{||}_{xy})}
{{\rho^{||}_{xx}}^2 + (|\bf{\rho^{||}}|\frac{\nu e^2}{h} + \rho^{||}_{xy})^2}
                      
\end{array}
\end{equation}    

Just as in single-layer Hall measurements of where expressions for the
longitudinal and transverse resistance are solved for the two unknowns,
density and Hall mobility, the above equations can be solved for the only
two unknowns in this problem, namely $\rho^{||}_{xx}$ and
$\rho^{||}_{xy}$, the components of the resistivity tensor in the parallel
layer.  We will demonstrate this analysis for data measured in the general
Van der Pauw geometry, although the same analysis can be adapted to Hall
bar and Corbino geometries.  All 4-point resistances were measured in an
Oxford $^3{\rm He}$ cryostat using standard lock-in techniques with a high
input impedance pre-amplifier ($R_{in}\sim 100~{\rm G}\Omega$) to avoid
measurement errors which might be mistaken for parallel conduction 
\cite{Fischer}.  

\begin{figure} 
\includegraphics[keepaspectratio,width=10cm]{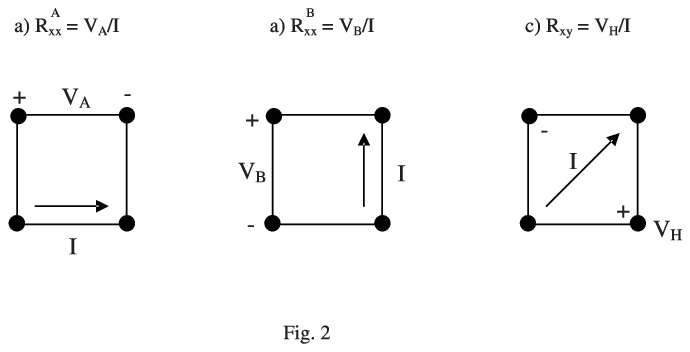}
\caption{ 
\label{VdP} 
The three measurements necessary to completely characterize the 
resistivity tensor at any magnetic field.} 
\end{figure}

The first sample of interest is a heterojunction quantum 
well with parallel conduction in the modulation doping layer.
We start experimentally by measuring the longitudinal resistance of the
sample at finite B-field in two complementary geometries $R^A_{xx}$ and 
$R^B_{xx}$ (see Fig. \ref{VdP}), and determine the total $B$-dependent sheet 
resistance $\rho^{tot}_{xx}(B)$ according to the Van der Pauw equations
\cite{vanderPauw}:

\begin{eqnarray}
\label{vdP}
\rho^{tot}_{xx} & = & \frac {\pi}{\rm{ln}(2)} f(x) \frac{R^A_{xx} 
+ R^B_{xx}}{2} \\
f(x) & = & 1 - \frac{{\rm ln}2}{2} \left( \frac{x-1} {x+1} \right)^2
- \left[ \frac{({\rm ln} 2)^2}{4} - \frac{({\rm ln} 2)^3}{12} \right]
\left( \frac{x-1} {x+1} \right)^4 \nonumber
\end{eqnarray}  

\noindent Here $x = R^A_{xx} / R^B_{xx}$ for $R^A_{xx} > R^B_{xx}$, and
$f(x) \leq 1$ is a weak function of $x$ of order unity.  The Hall
resistance $\rho^{tot}_{xy} = R_{xy}$ can be measured directly (Fig. 
\ref{VdP}).  

We now solve Eq. \ref{rhotot} for $\rho^{||}_{xx}$ and $\rho^{||}_{xy}$ in 
terms of the measured $\rho^{tot}_{xx}$ and $\rho^{tot}_{xy}$.  The 
total resistivity tensor $\bf{\rho^{tot}}$ inverted to give 
$\bf{\sigma^{tot}} = (\bf{\rho^{tot}})^{-1}$, and since $\bf{\sigma^{tot}}
= \bf{\sigma^{Q}} + \bf{\sigma^{||}}$,

\begin{equation}
\label{sigmatot}
\begin{array}{ccc}
\sigma^{tot}_{xx} = \sigma^{||}_{xx} = 
\frac {\rho^{tot}_{xx}}{|\rho^{tot}|},
&
&
\sigma^{tot}_{xy} = \sigma^{||}_{xy} + \frac{\nu e^2}{h} = 
\frac {\rho^{tot}_{xy}}{|\rho^{tot}|}
\end{array}
\end{equation}  

\noindent Knowing the filling factor index $\nu$ of a given quantum Hall
minimum, we solve for $\sigma^{||}_{xx} = \sigma^{tot}_{xx}$ and
$\sigma^{||}_{xy} = \sigma^{tot}_{xy} - \frac{\nu e^2}{h}$.  Inverting
$\bf{\sigma^{||}}$ to find $\bf{\rho^{||}} = (\bf{\sigma^{||}})^{-1}$
yields:

\begin{equation}
\label{parxxfin}
\rho^{||}_{xx} = \frac{\sigma^{||}_{xx}}{| \bf{\sigma^{||}}|} = \frac
{| \bf{\rho^{tot}}| \rho^{tot}_{xx}}
{( \rho^{tot}_{xx})^2 + ( \rho^{tot}_{xy} - | \bf{\rho^{tot}}| 
\frac { \nu e^2}{h})^2}
\end{equation}

\begin{equation}
\label{parxyfin}
\rho^{||}_{xy} = \frac{\sigma^{||}_{xy}}{|\bf{\sigma^{||}}|} = 
\frac {|\bf{\rho^{tot}}| (\rho^{tot}_{xy} - |\bf{\rho^{tot}}| 
\frac {\nu e^2}{h})}
{(\rho^{tot}_{xx})^2 + (\rho^{tot}_{xy} - |\bf{\rho^{tot}}| 
\frac {\nu e^2}{h})^2}
\end{equation}

\noindent which is nothing other than Eq. (\ref{rhotot}) solved for
$\rho^{||}_{xx}$ and $\rho^{||}_{xy}$. The result is plotted in Fig. 
\ref{delta_par} for values of $B$ where the 2DES Hall resistance is 
quantized.  

\begin{figure} 
\includegraphics{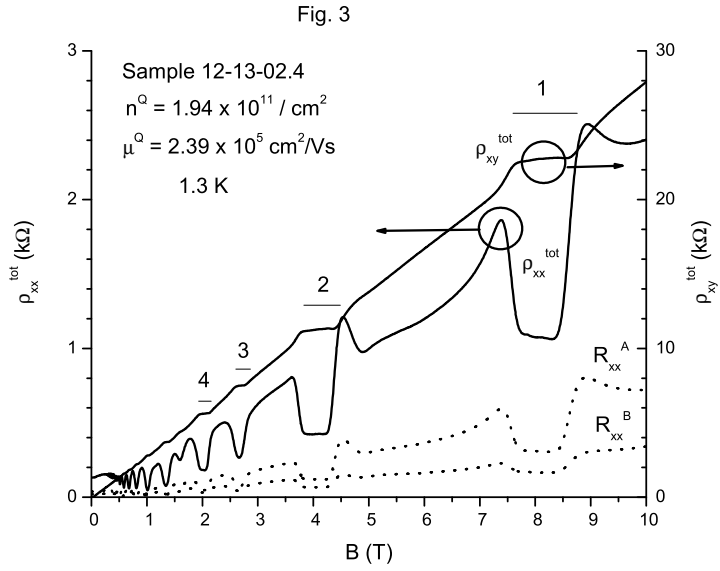}
\caption{
\label{delta_tot}
Plot of measured $\rho^{tot}_{xx}$, and $\rho^{tot}_{xy}$ for a
sample with parallel conduction in the modulation doped layer.  
$\rho^{tot}_{xx}$ is calculated from the Van der Pauw relation in
Eq.~\ref{vdP} using $R_{xx}^A$ and $R_{xx}^B$ as measured in the 
inset.}
\end{figure}

\begin{figure} 
\includegraphics{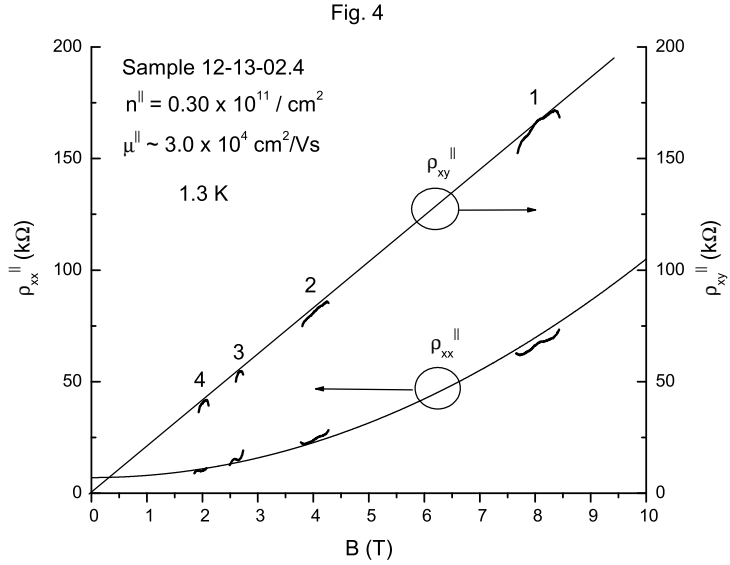}
\caption{
\label{delta_par}
Plot of $\rho^{||}_{xx}$ and $\rho^{||}_{xy}$ for the parallel 
conducting modulation doped layer calculated from the data in Fig. 
\ref{delta_tot} using Eqs. \ref{parxxfin} and \ref{parxyfin}.  The 
slope of $\rho^{||}_{xy}$ yields the carrier density in the parallel 
conductor $n^{||} = \frac{B}{\rho^{||}_{xy}} e$.}
\end{figure}

We can estimate the carrier density in the parallel layer with either 
of two methods.  First, the slope of $\rho^{||}_{xy}$ (both {\em within} 
each quantum Hall effect domain and across all filling factors) indicates 
a density of $n^{||} = \frac{B}{e \rho^{||}_{xy}} = 0.30 \times 10^{11} 
{\rm cm}^{-2}$.  Secondly, as a useful back-of-the-envelope shorthand, 
we note that in the limit of $\rho^{tot}_{xx} < \rho^{tot}_{xy}$ as in 
Fig. \ref{delta_tot}, Eq. (\ref{parxyfin}) can be simplified to,

\begin{equation}
\label{limitrhoxy}
\rho^{||}_{xy} = \frac{B}{n^{||}e} \approx 
\frac {\rho^{tot}_{xy}}
{1 - \rho^{tot}_{xy} \frac {\nu e^2}{h}}                     
\end{equation}  

\noindent Solving for $n^{||}$ gives 

\begin{equation}
\label{npar}
n^{||}  
\approx n^Q \left( \frac{ \frac{h}{\nu e^2} } {\rho^{tot}_{xy}} -1 \right)
\end{equation} 

\noindent where $n^Q = \frac{\nu e B(\nu)}{h}$ is the density of the QH
system, and $B(\nu)$ is the magnetic field value in the center of the
$\nu$-th plateau in $\rho^{tot}_{xy}$.  In Fig. \ref{delta_tot}, the flat 
lines above each $\rho^{tot}_{xy}$ plateau show the resistance quantum 
$\frac{h}{\nu e^2}$ demonstrating that the first term in parenthesis in 
Eq.~(\ref{npar}) is always greater than 1. The remarkable utility of 
Eq.~(\ref{npar}) is that the parallel carrier density can be estimated 
from a {\em single} $\rho^{tot}_{xy}$ {\em measurement}.  For example in 
$\rho^{tot}_{xy}$ from Fig. \ref{delta_tot}, the position of the $\nu = 1$ 
plateau at $B(\nu = 1) = 8.07~{\rm T}$ determines $n^Q = 1.95 \times 
10^{11}~{\rm cm}^{-2}$ and the plateau resistance of 
$\Delta\rho^{tot}_{xy} = 22.76 ~{\rm k}\Omega$ gives $n^{||} = 0.26 
\times 10^{11}{\rm cm}^{-2}$ -- within 15 \% of the more exact graphical 
analysis of Eq. (\ref{parxyfin}) and the resulting Fig. \ref{delta_par}.

\begin{figure} 
\includegraphics[]{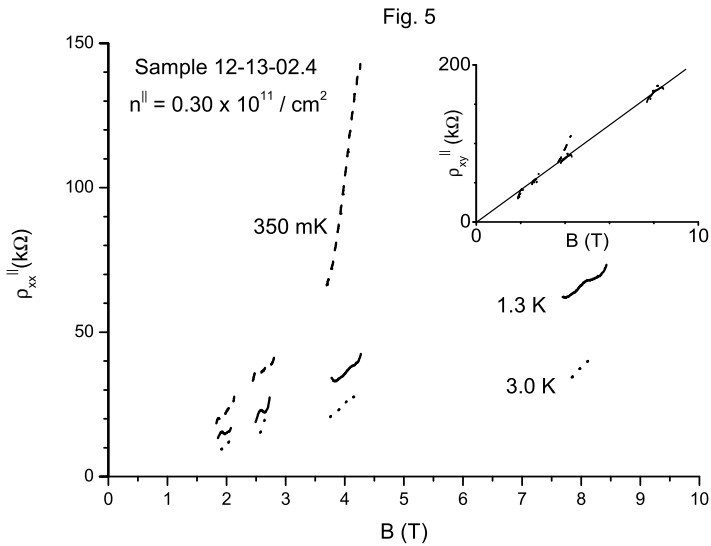}
\caption{
\label{freezeout}
Plot of $\rho^{||}_{xx}$ for the parallel conducting modulation doped 
layer at various temperatures.  The hopping conduction in this system is 
strongly temperature and magnetic field dependent. Inset: Plot of 
$\rho^{||}_{xy}$ for the parallel conducting layer.  Note that the 
density of hopping carriers is independent of $B$ and temperature, except above 4T 
at 350 mK under extreme magnetic freezeout.
}
\end{figure}

We can separately characterize the conductance of the parallel delta-doping 
layer as a result of this analysis.  Looking at the two components of the 
resistivity tensor in Fig. \ref{delta_par}, one observes first the sharp 
rise in $\rho^{||}_{xx}$ resulting from so-called magnetic freeze-out in 
this highly disordered delta-doping layer, expected in systems with hopping 
conduction.  The total number of mobile carriers in the delta layer, 
however, stays fixed as confirmed by the constant slope of $\rho^{||}_{xy}$.  
We can also make a rough estimate of the mobility of this layer, by 
extrapolating the parabolic fit shown in the figure to $B = 0$.  With an 
intercept resistivity of around $\rho^{||}_{xy} = 7 k\Omega$ and the known 
density, the $B = 0$ mobility of the parallel conduction layer is about 
$\mu^{||} = 3.0 \times 10^4 {\rm cm}^2/{\rm Vs}$.  In the temperature 
dependence shown in Fig. \ref{freezeout},
we see $\rho^{||}_{xx}$ increase with decreasing temperature, while $\rho^{||}
_{xy}$ remains fixed, indicating a temperature independent density.  Both 
behaviors are consistent with hopping conduction in the delta-layer.

\begin{figure} 
\includegraphics{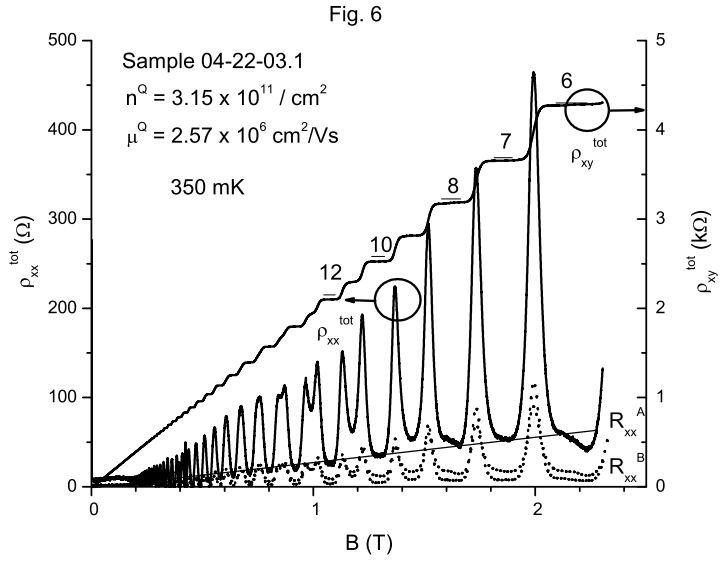}
\caption{
\label{2sub_tot}
Plot of $\rho^{tot}_{xx}$, $\rho^{tot}_{xy}$, $R_{xx}^A$ and
$R_{xx}^B$ for a quantum well with a lightly populated second subband.}
\end{figure}

\begin{figure}
\includegraphics[]{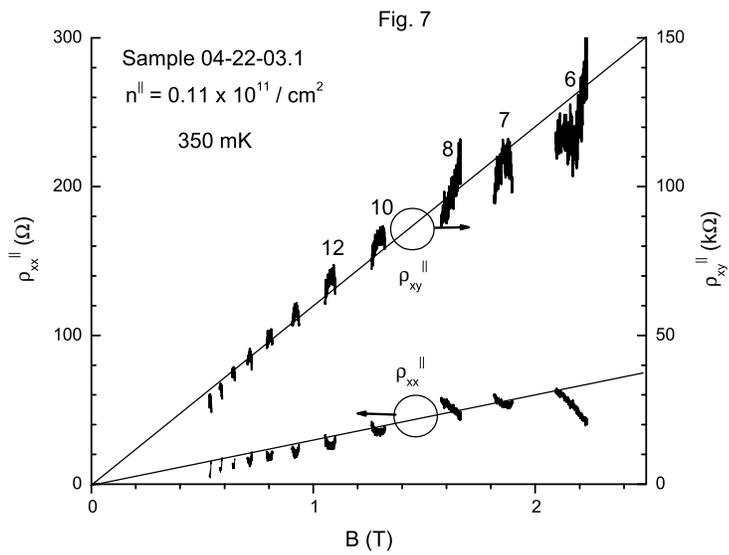}
\caption{
\label{2sub_par}
Plot of calculated $\rho^{||}_{xx}$ and $\rho^{||}_{xy}$ using the
data in Fig. \ref{2sub_tot} for a lightly populated second subband.}
\end{figure}

The same analysis can be applied to a two-subband quantum well sample.
In Fig. \ref{2sub_tot} we show the measured $\rho^{tot}_{xx}$ and
$\rho^{tot}_{xy}$ which are analized to give $\rho^{||}_{xx}$ and 
$\rho^{||}_{xy}$ as plotted in Fig. \ref{2sub_par}.  The measured parallel 
carrier density of $n^{||} = 0.11 \times 10^{11}~{\rm cm}^{-2}$ is in good 
agreement with self-consistent Poisson calculations which predict a second 
subband occupied to $n^{||}_{calc} = 0.15 \times 10^{11}~{\rm cm}^{-2}$.  

We note that the SdH signal from this low density second subband was too 
weak to give and estimable density using alternate the Fourier transform 
method.  Comparing the resistivity components in Fig. \ref{2sub_par} with 
Fig. \ref{delta_par}, one sees in $\rho^{||}_{xy}$ that once again the 
density of carriers is rather constant over the whole $B$-field range.  
The measured $\rho^{||}_{xx}$ is of order 50 $\Omega$ or less, implying a 
rather high mobility of over $10 \times 10^6 {\rm cm}^2/{\rm Vs}$ for 
this second subband.  Although this value is not quantitatively conclusive,
qualitatively it indicates that the low density subband which would normally 
have a correspondingly low mobility, benefits from the Thomas-Fermi 
screening of the densely populated subband to screen most of the disorder.

As a final note, we remind the reader that measurement errors arise if the 
input impedance of the lock-in $R_{in}$ is too low, and these pathological
signals look misleadingly like parallel conduction.  With a low $R_{in}$,
finite $R_{xx}$ minima of order $R_{min} = (h/\nu e^2)^2/R_{in}$ are
observed with the astonishing property that the false "parallel" signal
disappears upon reversing the current contact polarity.  Such subtleties
in quantum Hall measurement phenomena are discussed in detail elsewhere
\cite{Fischer}.

In conclusion, we have demonstrated a new analysis for the resistivity of
parallel 2D systems that determines the resistivity tensor of one layer at
magnetic fields where the other layer is in the quantum Hall regime.  We
demonstrate the validity of this analysis for the case of a quantum well
with two occupied subbands, as well as for double-quantum wells. We also
demonstrate the analysis of carrier density in parallel conducting
modulation doped layers.  In the latter case, this technique will be
particularly useful for crystal growers in characterizing doping
efficiency, because it elimates the expensive and time-consuming
trial-and-error method currently used for doping calibration of parallel
conduction layers.

{\it Acknowledgement --} This work was supported in part by the DFG (SFB
348).  M.G. thanks the A.v. Humboldt Foundation for support.

$(a)$ Email: mgrayson@alumni.princeton.edu

\clearpage
\begin{figure}[] 
\includegraphics[]{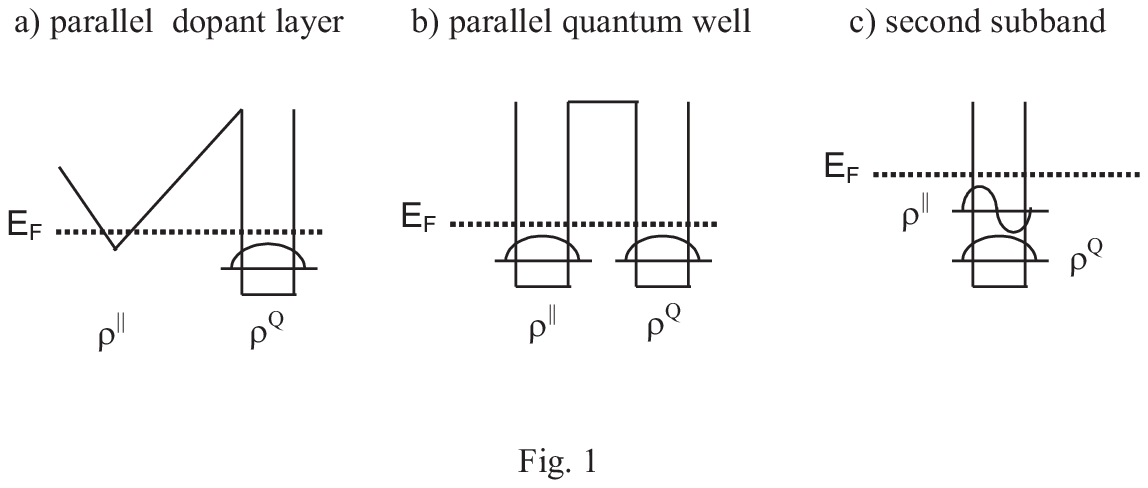}
\caption{ 
\label{systems} 
Conduction band schematic for the
three kinds of parallel conduction systems of interest: a) the dopant
layer of a modulation-doped heterostructure, b) a double-quantum well
system, or c) a quantum well with a second occupied subband.} 
\end{figure}
\clearpage

\end{document}